\def\etal     {{\it et al. \/}}

\def\eg       {{\it e.g. \/}}
\def\ie       {{\it i.e. \/}}

\newcommand{\ee}[1]{\cdot10^{#1}}
\newcommand{\mr}[1]{\mathrm{#1}}
\newcommand{\unit}[1]{\,\mathrm{#1}}

\newcommand{\us}{\,\mu{\rm s}}

\newcommand{\Hzf}{H_\mr{zf}}
\newcommand{\Hlocal}{H_\mr{magn}}
\newcommand{\Hother}{H_\mr{other}}

\newcommand{\Brms}{B_\mr{rms}}
\newcommand{\DD}{\bf \mathcal{D}}
\newcommand{\dw}{\delta\omega}
\newcommand{\Dw}{\Delta\omega}
\newcommand{\DB}{\Delta B}
\newcommand{\Dth}{\Delta\theta}
\newcommand{\ye}{\gamma_\mr{e}}
\newcommand{\yn}{\gamma_\mr{n}}
\newcommand{\vecr}{{\bf{r}}}
\newcommand{\vecnr}{{\bf{\hat{r}}}}
\newcommand{\ez}{{\bf{\hat{z}}}}

\documentclass[prl,aps,twocolumn]{revtex4}

\usepackage{graphicx}
\usepackage{bm}
\usepackage{amsmath}
\usepackage{amssymb}
\usepackage{rotating}
\usepackage[colorlinks=true, pdfstartview=FitV, linkcolor=blue, citecolor=blue, urlcolor=blue]{hyperref} 

\begin{document}

\global\emergencystretch = .1\hsize 

\title{Scanning magnetic field microscope with a diamond single-spin sensor}

\author{C. L. Degen$^{1}$}
  \email{degenc@gmail.com} 
  \affiliation{
   $^1$IBM Research Division, Almaden Research Center, 650 Harry Road, San Jose, CA 95120, USA.}
\date{\today}

\begin{abstract}
We describe a scanning device where a single spin is used as an ultrasensitive, nanoscale magnetic field sensor. As this ``probe spin'' we consider a single nitrogen-vacancy defect center in a diamond nanocrystal, attached to the tip of the scanning device. Changes in the local field seen by the probe spin are detected by optically monitoring its electron paramagnetic resonance transition. The room-temperature scanning device may be useful for performing nanoscale magnetic resonance imaging and spectroscopy, and for the characterization of magnetic nanostructures down to the single atom level.
[DOI: \href{http://dx.doi.org/10.1063/1.2943282}{10.1063/1.2943282}]
\end{abstract}

\maketitle

Investigating magnetism at the nano- and atomic scale is a key issue both for understanding fundamental physical properties of matter
and as the enabling ingredient for magnetism-based data storage and spintronic devices.
Tools for studying and imaging magnetic structures with nanometer resolution are, for example,
polarization sensitive electron microscopy (SEMPA), magnetic force microscopy (MFM), magnetic resonance force microscopy (MRFM),
and scanning tunneling microscopy (STM) \cite{hopster03}.
Some of them (\eg, STM) even extend down to single magnetic atoms \cite{heinze00} while others (\eg, MRFM)
allow observing the faint magnetism of single electron or nanoscale volumes of nuclear spins \cite{rugar04,mamin07}.

\begin{figure}[b!]
      \begin{center}
      \includegraphics[width=0.48\textwidth]{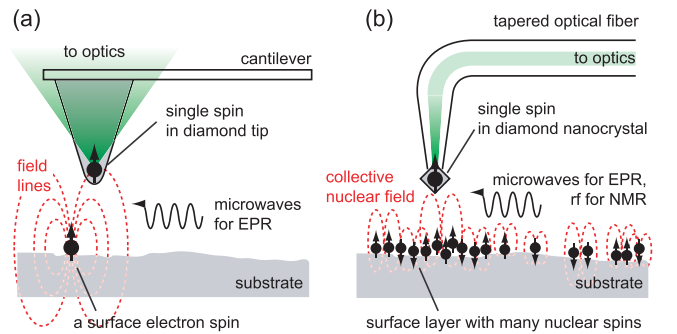}
      \end{center}
      \caption{Diamond-based scanning spin microscope. The single spin associated with a N-V defect in diamond serves as an ultrasensitive magnetometer with nanoscale spatial resolution. Optical monitoring of the change in the spin's EPR frequency reveals magnetic coupling to, for example, a single surface spin (a) or an ensemble of nuclear spins (b). Two experimental implementations are suggested, based on (a) AFM and (b) scanning near field optical microscopy.}
      \label{fig1}
\end{figure}

Here we describe an alternative scanning device integrating a single spin as a sensitive and high resolution
magnetic field sensor that may be a useful addition to the toolbox of nanoscale magnetic probes.
The idea behind our approach,
which is based on the ``spin microscope'' proposed by Chernobrod and Berman \cite{chernobrod05} and Berman \etal \cite{berman06},
is illustrated in Fig. \ref{fig1}:
Attached to the tip of a scanning device is a ``probe spin'' whose state can be monitored using optically detected magnetic resonance.
In our case, this probe spin will be a single nitrogen-vacancy (N-V) defect in the diamond tip of an atomic force microscope cantilever (Fig. \ref{fig1}(a)).
If the probe spin is brought near a substrate, it will feel the presence of any local magnetic fields
emanating from the surface, causing a shift of its electron spin resonance (EPR) frequency.
This shift can then be detected, \eg, by exciting the EPR transition with a microwave field
and monitoring the change in photoluminescence of the probe spin.

The magnetic field $B$ responsible for the frequency shift can have a variety of origins,
which is one reason why we believe that the proposed concept is very powerful.
We will be particularly interested in the situations where $B$ is the static field of a magnetic nanostructure,
the magnetic dipole field of a single magnetic atom or an isolated electron spin, or the collective magnetic field of an ensemble of nuclear spins.

The probe spin considered in this Letter is the single spin associated
with the N-V defect center in diamond \cite{jelezko06}.
Not only is this defect one of the few solid-state systems where the spin state can be directly measured,
but it also combines a line of extraordinary properties that make it very attractive for such a scanning device.
In particular, these are excellent chemical- and photostability, extraordinarily long spin lifetimes,
and the fact that single-spin detection is possible under ambient conditions \cite{jelezko06}.

%

%
\begin{table*}[htb]
\caption{Example parameters for prospective applications.}
\begin{tabular*}{\textwidth}{llll}
\hline\hline
																	&											& Scheme A (``dc-type''):		& Scheme B (``ac-type''):		\\
\hline
Quantity &  & Direct shift of EPR resonance & Modulation/Attenuation of spin echo \\
1. Resolvable frequency shift				& $\dw/2\pi \footnotemark[1]$  & typ. $5\unit{MHz}$ \cite{popa04,epstein05}						& typ. $2\unit{kHz}$ 					\\
2. Associated spin lifetime					&	$\tau=\dw^{-1}$\quad\quad				& typ. $30\unit{ns}$ 									& typ. $100\unit{\us}$ \cite{kennedy03,jelezko06}	\\

Magnetometry  \\
3. Minimal resolvable field, low $B$ \footnotemark[2] 				& $\DB$  							
 & $0.2\unit{mT} \times (\cos\theta)^{-1}$					& $60\unit{nT} \times (\cos\theta)^{-1}$ \\
4. Minimal resolvable field, high $B$ \footnotemark[2] 				& $\DB$
 & $0.2\unit{mT}$ 														& $60\unit{nT}$ \\
5. Minimal resolvable field angle, low $B$ \footnotemark[2]   & $\Dth$
 & $2\ee{-1}\unit{rad} \times 1\unit{mT}/B \times (\sin\theta)^{-1}$ 	& $6\ee{-5}\unit{rad} \times 1\unit{mT}/B \times (\sin\theta)^{-1}$ \\
6. Minimal resolvable field angle, high $B$ \footnotemark[2] & $\Dth$
 & $1\ee{-3}\unit{rad} \times (\sin2\theta)^{-1}$		& $4\ee{-7}\unit{rad} \times (\sin2\theta)^{-1}$ \\

Single spin detection \\
7. Separation for detecting single electron spin 				& $r$ 								&	$<3\unit{nm}$			 								& $<40\unit{nm}$ \\
8. Separation for detecting single proton spin 					& $r$									& ($<0.3\unit{nm}$)									& $<5\unit{nm}$ \\

Nuclear magnetic resonance imaging \\
9. Fluctuating nuclear field,\\
   $10\unit{nm}$ above proton-rich surface (see text)        & $\DB_\mr{rms}$  		& \multicolumn{2}{c}{$0.2\unit{\mu T_\mr{rms}}$}\\
\hline\hline
\label{table:examples}
\vspace{-0.4cm}
\footnotetext[1]{$\dw$ is $0.5\times$ the resonance linewidth (full width at half maximum).}
\footnotetext[2]{``Low $B$'' and ``high $B$'' as referred to $D/\ye\approx100\unit{mT}$.}
\end{tabular*}
\end{table*}
%

In a quantum mechanical analysis, we may describe the probe spin ${\bf S}$ by the following spin Hamiltonian \cite{morrish65},
\begin{equation}
H = \Hzf + \Hlocal + \Hother.
\end{equation}
Here, $\Hzf = {\bf S \cdot\DD\cdot S}$ describes the zero-field splitting of the electronic ground state,
where $\DD$ is an (axially symmetric) tensor with splitting constant $D=2\pi\times 2.88\unit{GHz}$
that separates the $m_S=0$ from the (degenerate) $m_S=\pm1$ sublevels \cite{jelezko06}.
$\DD$ is oriented along the N-V symmetry axis (a $\langle111\rangle$ crystal axis),
which we assume to be the $\ez$ axis. 
$\Hlocal = \hbar\ye{\bf B\cdot S}$, where $\ye$ is the electron gyromagnetic ratio,
comprises all local magnetic fields ${\bf B}$ seen by the probe spin that will be of further interest.
$\Hother$ includes all other interactions, such as hyperfine couplings, dipolar couplings to distant defects,
optical fields, or strain on the crystal breaking the symmetry of $\DD$.
These interactions may even be substantial (and may form a ``nuisance'' for practical implementation
by making the spectrum complicated or affecting spin relaxation \cite{popa04}),
but are not of immediate interest for the purpose of this Letter.

$\Hzf$ and $\Hlocal$ are competing terms and depending on the strength of ${\bf B}$,
one or the other will dominate the Hamiltonian. In the situation where $B\equiv|{\bf B}|$ is weak,
$B\ll D/\ye$ (about $100\unit{mT}$), $\Hlocal=\hbar\ye B S_z\cos\theta$ is a perturbation
to $\Hzf = \hbar D[S_z^2-\frac{1}{3}S(S+1)]$.
The shift $\Dw$ of the spin resonance frequency due to a change in magnitude $\DB$ or direction $\Dth$
of the magnetic field is then approximately given by
$\Dw = \ye \cos(\theta) \DB  - \ye B \sin(\theta) \Delta\theta$,
where $\theta$ is the angle between the principle axis of $\DD$ (the $\ez$ axis) and ${\bf B}$.
For strong fields $B\gg D/\ye$, the Zeeman term $\Hlocal=\hbar\ye B S_{z'}$ sets the spin's quantization axis ($\ez'||{\bf B}$)
and $\Hzf = \hbar D[S_{z'}^2-\frac{1}{3}S(S+1)] \, \frac{1}{2}(3\cos^2\theta-1)$ can be treated as a perturbation.
The frequency shift is then about $\Dw = \ye \DB  - \frac{3}{2}D\sin(2\theta) \Delta\theta$.
In the intermediate range ($\ye B \approx D$) variations in magnetic field
certainly lead to EPR frequency shifts in general, however, they might be harder to interpret
and the crossover of spin energy levels may mask optical transition rates \cite{epstein05}.

What would be an effective way to detect the frequency shift, and what will be the magnetic sensitivity of the device?
In order to directly resolve the shift in the EPR spectrum,
$\Dw$ must be of the order of the resonance linewidth or larger.
For diamond, the natural EPR linewidth is typically a few MHz \cite{popa04,epstein05},
corresponding to a few Gauss of Zeeman field. A representative set of parameters is summarized in Table I (lines A1-A6). 

This ``natural linewidth'', however, is often inhomogeneously broadened,
and one can potentially do much better by observing spin precession in a spin echo-type experiment.
(This was demonstrated, for example, in measurements of the Stark effect of small electric fields \cite{vanoort90}.)
The minimal detectable frequency shift is then on the order of the inverse of the $T_2$ time.
Since $T_2$ in diamond can be exceptionally long [exceeding $100\unit{\us}$ (Refs. \cite{jelezko06} and \cite{kennedy03})],
very small field changes will be measurable. If we associate a ``linewidth'' $\dw=(T_2)^{-1}$ with $T_2$,
we find that field changes as small as a few tens of nanotesla can be resolved (lines B1-6 in Table I).

It is important to notice that modulation (or attenuation) of the spin echo will only occur for time-dependent (ac) magnetic fields 
that fluctuate on the timescale of $T_2$. Static fields will be refocused by a spin echo,
while very rapidly oscillating fields will not lead to significant dephasing \cite{slichter}.
This may seem restrictive, but it may not be for real experiments. Static fields, such as those originating from
a (ferro)magnetic nanostructure, are often substantial and should be easily detectable as a direct line shift.
Weak magnetic fields, such as those present close to single electron or small numbers of nuclear spins,
are often fluctuating, or can be deliberately made to do so, for example using magnetic resonance pulses
or also by rapidly moving (vibrating) the tip.

At this point it is convenient to review how the instrument may be implemented.
One possible design, sketched in Fig. \ref{fig1}(a), combines an atomic force microscope (AFM) cantilever with a diamond tip
and a confocal optical microscope \cite{vanhulst92}. Another possibility is the attachment or direct growth
of a diamond nanocrystal on the end of a bent and tapered optical fiber (Fig. \ref{fig1}(b)),
where the fiber serves the dual purpose of optical waveguide and scanning element \cite{muramatsu98,kuhn01}.
Both approaches
have the added advantage of employing the cantilever, or fiber,
as force sensors in the traditional manner of force microscopy.
This would allow one to simultaneously obtain the surface topography,
which may also be helpful for navigating the probe.

The separation between the N-V center in the tip and the substrate will typically be a few nanometers.
Placing a N-V defect that close to the tip --- without disturbing its exceptional optical and spin properties ---
is not only a substantial technical challenge but also presumes that the defect is stable within nanometers from the crystal surface.
Demonstrated minimum sizes for nanocrystals with functional N-V centers are as small as $15\unit{nm}$ \cite{rabeau07,sonnefraud08},
less than an order of magnitude away from the lengthscale of the most promising applications summarized in Table I. 

We notice that such a scanning device, should it succeed in operating at the nanometer level,
may have a considerable range of applications. 
In the following we point out three specific examples of how the instrument might be used.

As a first example we consider the imaging of a (ferro)magnetic nanostructure and the study of its magnetic properties.
What features can be discerned?
It is well known from magnetic recoding that spatial frequencies in the field caused
by the fine structure of typical length $\sim L$ of a magnetic substrate
will decay exponentially $\propto e^{-r\pi/L}$ with distance $r$ from the surface \cite{white84}.
Hence, a probe spin scanning at separation $r$ from the surface will have a lateral resolution of typically $\pi r$.
Since the n-v spin is an excellent field sensor, however, 
magnetic characteristics (such as a magnetization curve) of much smaller features
may still be studied, provided they are isolated enough.

This sensitivity may well extend to single atoms.
As a second example we hence consider the detection of the magnetic dipole field of a single electron spin (as suggested in Fig. \ref{fig1}(a)).
The interaction of the probe spin ${\bf S}$
with a surface spin ${\bf S'}$ may be described by a dipolar Hamiltonian \cite{slichter},
\begin{equation}
\Hlocal = \frac{\mu_0}{4\pi}\frac{\hbar^2\ye\ye'}{r^3} \left\{ {\bf S\cdot S'} - 3({\bf S\cdot\vecnr})({\bf S'\cdot\vecnr}) \right\}
\label{eq:dipolar}
\end{equation}
where $r\equiv|\vecr|$ is the spatial separation and $\vecnr\equiv\vecr/|\vecr|$ is the normalized interspin vector.
Assuming that ${\bf S}$ and ${\bf S'}$ are aligned,
the $z$-component of the dipole field seen by the probe spin (or vice versa, the surface spin)
is $B = \frac{\mu_0}{4\pi}\frac{\hbar\ye'}{r^3} [1-3\cos^2(\theta)] S'_z$, where $\theta$ is the angle between $\vecnr$ and $\ez$
and $S'_z$ is the state of the surface spin.
We can solve the above expression for $r$ in order to find out what proximity is needed to create a detectable frequency shift.
In a configuration where the probe sits right over the surface spin we find that $r$ is well above $10\unit{nm}$ (Table I, line 7).

It may also be possible to read out the state $S'_z$ of the surface spin. This could, for example, be done by transferring
$S'_z$ to the probe spin state $S_z$ using spin-echo double resonance \cite{slichter96double} and then measuring $S_z$ optically \cite{jelezko02}.
Requirement is that the surface spin lifetime $T_1$ is sufficiently long to allow for state transfer,
\ie, $T_1>(\ye B)^{-1}$, which will typically be in the microsecond range.

Similar estimations can be made for the detection of a single nuclear (proton) spin.
Since the proton moment is roughly $1\,000\times$ weaker than the electron moment,
about $10\times$ closer proximity is necessary, \ie at most a few nanometers (line 8 in Table I).
Such close separations will be very challenging to realize, but they are not out of the question.

The collective dipole field of a large number of nuclei, on the other hand, might be readily observable [Fig. \ref{fig1}(b)].
As a third example, we consider the situation where the probe spin is positioned over a homogeneous surface layer
of material containing many nuclear spins. Specifically, we assume that the vertical separation is $r = 10\unit{nm}$
and that the substrate has a proton spin density of $\rho=5\ee{28}\unit{spins/m^3}$, as typical for an organic material.
The protons on the surface will give rise to a fluctuating statistical spin polarization producing
a collective field $B \propto \mu_\mr{p}\sqrt{N}$,
where $\mu_p$ is the proton magnetic moment and $N$ is the number of spins in the vicinity of the tip \cite{degen07}.
The rms value of $B$ can be calculated explicitly by integration over the individual nuclear dipoles,
$B_\mr{rms} = \frac{\mu_0}{4\pi}\frac{\hbar\yn}{2}
\left\{ \int d^3\vecr \rho(\vecr) [1-3\cos^2\theta(\vecr)]^2|\vecr|^{-6} \right\}^\frac{1}{2}$,
where $\yn$ is the nuclear gyromagnetic ratio and $\theta$ the angle between $\vecr$ and $\ez$.
For our example, we find that $\Brms\sim 0.2\unit{\mu T}$ (Table I, line 9) --- well above our detection limit.
This $\Brms$ is equivalent to the field of about $80\,\mu_{\mr p}$ situated $10\unit{nm}$ right below the probe.

Similar to other nanoscale magnetic resonance imaging techniques (such as MRFM),
the scanning device could be combined with a nanoscale magnetic tip in order to improve the spatial resolution to well below $r$ \cite{mamin07}.
The elemental selectivity of nuclear magnetic resonance could also be used to discriminate various chemical species.
Finally, magnetic resonance will be valuable to distinguish nuclear (or electron) dipole fields from other local magnetic fields
influencing the probe spin resonance.

The device described herein might eventually allow the imaging
of biological structures and organic surface layers, and was in fact motivated by these ideas.
Even at a probe-to-sample distance of $10\unit{nm}$, its sensitivity would outperform MRFM,
currently the most sensitive magnetic resonance detection technique and able to detect about $100$ proton moments \cite{mamin07},
by at least an order of magnitude.
Unlike MRFM, however, a diamond-based magnetic field sensor is compatible with room temperature operation and
might even enable the study of biological systems under physiological conditions.

Helpful discussions with Martino Poggio, John Mamin, and Dan Rugar are gratefully acknowledged.
The material presented in this manuscript was originally part of a research proposal
by the author submitted in October of 2007 \cite{degen07prop}.
While finishing this manuscript, we learned about
a similar idea put forward by Lukin \cite{lukin08}.

\end{document}